\documentclass[aps,prb,twocolumn,superscriptaddress]{revtex4}
\newcommand{\absz}[1]{\mbox{$\mid\!\!#1\!\!\mid$}}
\usepackage{graphicx}
\usepackage{amsmath}

\begin{document}

\title{A simple model for the vibrational modes in honeycomb lattices}
\author{J{\'o}zsef Cserti}
\email{cserti@galahad.elte.hu}
\affiliation{Department of Physics of Complex Systems, E{\"o}tv{\"o}s
University}
\author{G{\'e}za Tichy}
\affiliation{Department of Solid State Physics, E{\"o}tv{\"o}s
University, 
H--1117 Budapest, P\'azm\'any P{\'e}ter s{\'e}t\'any 1/A,
Hungary
}


\begin{abstract}
The classical lattice dynamics of honeycomb
lattices is studied in the harmonic approximation.
Interactions between nearest neighbors are represented by springs
connecting them.
A short and necessary introduction of the lattice structure
is presented.
The dynamical matrix of the vibrational modes is then derived,
and its eigenvalue problem is solved analytically.
The solution may provide deeper insight into the nature of the
vibrational modes.
Numerical results for the vibrational frequencies are presented.
To show that how effective our method used for the case of honeycomb 
lattice is, we also apply it to triangular and square lattice structures. 
A few suggested problems are listed in the concluding section.
\end{abstract}


\maketitle

\section{Introduction}
In crystals, atoms vibrate about their equilibrium position.
Calculation of the vibrational frequencies and modes is an
important and presently actively studied subject in solid-state
physics. To interpret various properties of crystal lattices---for
example, specific heat, thermal expansion coefficients, elastic
constants---it is essential to take into account the lattice
vibrations. Fortunately, several good textbooks of the subject are
available (here we give just a small selection of the vast
literature \cite{Kittel, Ziman,Ashcroft,Harrison,Jones}), in which
the quantum and classical treatments of lattice dynamics are
presented lucidly. The vibration of the atoms depends 
on the interatomic interaction within the
crystal. The interatomic interaction potential is
often approximated by including only quadratic terms of the
displacement of the atoms. This is called the \emph{harmonic
approximation}, and is the usual starting point for developing the
theory of lattice dynamics. 
In the framework of classical mechanics, lattice dynamics can be treated 
through the equations of motion for the atoms. 
To determine the vibrational frequencies and the corresponding modes one
needs to calculate the eigenvalues and the eigenvectors of the so-called
dynamical matrix, which can be obtained from the interatomic interaction
potential. If the dynamical matrix is known, the eigenvalue
problem is straightforward, though only numerical solutions are available
in many cases. 
Numerous
texts\cite{Kittel,Ziman,Ashcroft,Harrison,Jones,Goldsmid,Mihaly_Laci}
provide problems for the calculation of the vibrational modes 
in different crystal lattices.
Fortunately, there are some well-known examples---such
as linear chains with/without bases, square and cubic lattices---in
which analytical methods can be used
\cite{Kittel,Ziman,Ashcroft,Harrison,Jones,Goldsmid,Mihaly_Laci}. 
These examples are important for a deeper understanding of the theory of 
lattice dynamics.

\begin{figure}[hbt]
\includegraphics[scale=0.6]{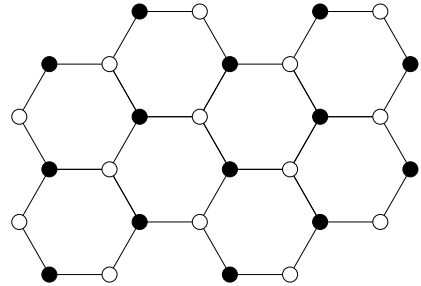}
\caption{Honeycomb lattice with two different atoms (black and white circles).
The straight lines correspond to unstrained springs of force constant
$D$ connecting the nearest neighbor atoms in equilibrium state.
The masses of the black and white atoms are $m_1$ and $m_2$.
\label{honey:fig}}
\end{figure}

In this paper we shall give another nontrivial example, namely the lattice
dynamics of the honeycomb structure shown in Fig.\ \ref{honey:fig}. 
The classical lattice dynamics of the honeycomb
lattice is modelled by using the harmonic approximation 
and assuming only nearest neighbor interactions between the atoms.
In this simplest model the interaction between nearest neighbors 
can be represented by springs connecting them.
It is assumed that black and white atoms have different masses, $m_1$
and $m_2$, respectively. 
The problem can be solved analytically and its solutions may provide
insight into the nature of the vibrational modes.
As it will be seen below, the analytical method is possible
because of the symmetry properties of the honeycomb lattice.
As a pre-requisite to the solution, a short and necessary introduction
of the lattice structure is presented.
More details on crystal structures can be found in many books
on solid-state physics\cite{Kittel,Ziman,Ashcroft,Harrison,Jones,Lubensky}.
Finally, in connection with the honeycomb lattice,
the intensive research on nanotubes\cite{nanotube-book} 
should be mentioned.
Nanotubes are quasi-one-dimensional cylindrical structures
in which a two-dimensional honeycomb lattice of crystalline graphite
is rolled up into a cylinder. The work presented in
this paper may serve as an introductory study of the far
more complex lattice dynamics of graphite and nanotubes.

The general method developed for calculating the vibration modes of
honeycomb lattices proves to be useful for studying other lattice
structures. Therefore, we shall present a few numerical results 
for the vibrational frequencies of equilateral triangular lattices 
and square lattices in which both the nearest neighbor
and second nearest neighbor interactions are taken into account.   

Our approach used in this paper may provide a didactically 
useful examples for introducing the basic lattice dynamical notions 
developed in solid-state physics.

The rest of the text is organized as follows. In Sec.\ \ref{force:seq}
a formula for the spring force acting on the atoms is derived.
In Sec.\ \ref{motion:seq} the description of the honeycomb lattice and the
equations of motion for the atoms are given.
In Sec.\ \ref{algebra:seq} our analytic method for solving
the eigenvalue problem is presented.
In Sec.\ \ref{res:seq} some numerical results are mentioned.
A few results are presented in Sec.~\ref{3szog4szog:sec} 
for the case of triangular and square lattices.
The conclusions and some suggested problems are given in Sec.\ \ref{veg}.

\section{The Force}
\label{force:seq}

First, the force acting on an atom by a spring connecting the
atoms is calculated.
The springs are assumed to be ideal, ie.\ the spring force
is proportional to the change in the spring's length.
Denoting the vector between the two endpoints of the unstrained (strained)
spring by ${\bf d}$ (${\bf r}$), the change in the length of the spring is
$\absz{{\bf r}}-\absz{{\bf d}}$. If the spring constant of the ideal
springs is $D$ then the restoring force is
\begin{equation}
{\bf F}= -D\left(\absz{{\bf r}}-\absz{{\bf d}} \right)
\frac{{\bf r}}{\absz{{\bf r}}}.
\end{equation}
It is useful to express the force in terms of the displacement vector
${\bf u}={\bf r}-{\bf d}$.
In the harmonic approximation the strain of the spring is assumed to
be small, ie. second order terms are neglected in ${\bf u}$. Explicitely:
$\absz{{\bf u}}/d=\absz{{\bf r}-{\bf d}}/d \ll 1$.
Thus, we have
$\absz{{\bf r}} = d\sqrt{1+2\frac{{\bf du}}{d^2}+\frac{{\bf u}^2}{d^2}}
\approx d\left(1+\frac{\bf du}{d^2} \right)$ and
the force can be written as
\begin{equation}
{\bf F} =  -D\, \frac{\left({\bf du}\right){\bf d}}{d^2}.
\label{force:def}
\end{equation}
It is convenient to define the outer or direct
product ${\bf a}\otimes{\bf b}$ of two vectors ${\bf a}$ and
${\bf b}$\cite{numrec,Arfken}.
This is a matrix whose $\alpha,\beta$
element is a product of the $\alpha$th component of ${\bf a}$ and
the $\beta$th component of ${\bf b}$, that is
\begin{equation}
{\left({\bf a}\otimes{\bf b}\right)}_{\alpha \beta}=a_\alpha^* b_\beta,
\label{diadic_prod}
\end{equation}
where ${}^*$ denotes the complex conjugation (of complex vectors).
The following direct-product identities will prove useful in the
calculations below:
\begin{subequations}
\begin{eqnarray}
({\bf a}\otimes{\bf b})\,{\bf c} &=& {\bf a}\, ({\bf b} {\bf c}),
\label{direct-1:id}  \\
({\bf a}\otimes{\bf b}) ({\bf c}\otimes{\bf d}) &=&
({\bf b} {\bf c}) \, {\bf a}\otimes{\bf d}.
\label{direct-2:id}
\end{eqnarray}
\end{subequations}
In direct-product notation, the (\ref{force:def}) force can be rewritten
as
\begin{equation}
{\bf F} =
-D\, \frac{{\bf d}\otimes{\bf d}}{d^2} \, {\bf u}.
\label{force_diadic_form}
\end{equation}
From this form it is clearly seen that the force in the harmonic approximation
is proportional to the displacement vector ${\bf u}$, and that the constant of
proportionality is the product of the spring constant $D$ and the direct
product of the unit vector ${\bf d}/|{\bf d}|$ with itself.
In the following section the equations of motion of the atoms are
formulated by using the above form of the spring force acting on the atoms.

\section{The Equations of Motion}
\label{motion:seq}

\begin{figure}[hbt]
\includegraphics[scale=1.0]{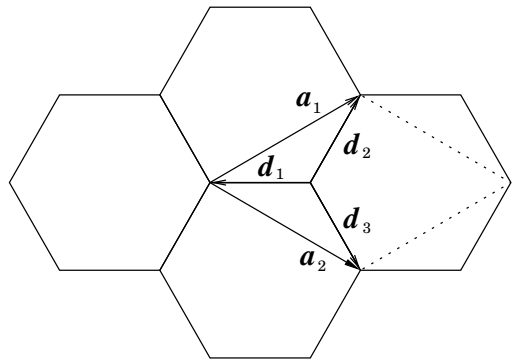}
\caption{The primitive translation vectors of the honeycomb lattice are
vectors ${\bf a}_1$ and ${\bf a}_2$; the primitive cell is a
parallelogram whose sides are these two vectors and the two dotted lines.
The three vectors ${\bf d}_1,{\bf d}_2,{\bf d}_3$ will be useful in the
calculations below.
\label{honeyS:fig}}
\end{figure}

In Fig.~\ref{honey:fig} the position vector ${\bf R}$ of the black
atoms in the honeycomb structure can be given in the form
\begin{equation}
{\bf R}=j_1{\bf a}_1 +j_2{\bf a}_2,
\label{d-lattice}
\end{equation}
where ${\bf a}_1$ and ${\bf a}_2$ are the independent primitive
translation vectors shown in Fig.~\ref{honeyS:fig}, while $j_1$ and $j_2$
range through all integer values (i.e.\ positive and negative integers, as
well as zero).
The primitive translation vectors have the same magnitude
$|{\bf a}_1|=|{\bf a}_2|= \sqrt{3}a$, where $a=\left|{\bf d}_1 \right|$ is
the length of hexagons' side. The primitive cell of the honeycomb lattice
can be chosen as the parallelogram whose non-parallel sides are the two
primitive translation vectors ${\bf a}_1$ and ${\bf a}_2$
(in Fig.~\ref{honeyS:fig} the sides of the parallelogram are
${\bf a}_1$, ${\bf a}_2$ and the two dotted lines).
In the honeycomb lattice each primitive cell with a {\em two-point basis}
contains two atoms (black and white circles in Fig.~\ref{honey:fig});
one is at the common starting point of vectors ${\bf a}_1$ and ${\bf a}_2$,
while the other one is at $1/3({\bf a}_1+{\bf a}_2)$ in Fig.~\ref{honeyS:fig}.

It is useful to introduce three unit vectors (${\bf n}_1,{\bf n}_2,{\bf n}_3$) by
\begin{equation}
{\bf n}_i=\frac {{\bf d}_i}{|{\bf d}_i|} \mbox{ , } i=1,2,3.
\label{BS:18:1a}
\end{equation}
The vectors ${\bf d}_1,{\bf d}_2,{\bf d}_3,$ are shown in Fig.
\ref{honeyS:fig}; their magnitude is $|{\bf d}_i| = a$.
Thus, the primitive translation vectors are
${\bf a}_1= {\bf d}_2 - {\bf d}_1 = a({\bf n}_2 - {\bf n}_1)$
and ${\bf a}_2= {\bf d}_3 - {\bf d}_1 = a({\bf n}_3 - {\bf n}_1)$.
Later on the following identities for vector ${\bf n}$
will frequently be used:
\begin{subequations}
\begin{eqnarray}
\sum_{j=1}^3  {\bf n}_j &=& 0,
\label{sum-n:id}  \\
\sum_{j=1}^3  {\bf n}_j \otimes {\bf n}_j &=& \frac {3}{2} \, {\bf E},
\label{n_jon_k:id}  \\
({\bf n}_j{\bf n}_k ) &=& \begin{cases}
\phantom{-}1 & \text{if $j=k$},   \\[1ex]
-\frac{1}{2} & \text{if $j \ne k$,}
\end{cases}
\label{n_jn_k:id}
\end{eqnarray}
\label{diad:id}
\end{subequations}

and ${\bf E}$ is the $2 \times 2$ unit matrix.

The displacements of the two atoms in the primitive cell
specified by the lattice vector ${\bf R}$
are denoted by ${\bf u}({\bf R})$ and ${\bf v}({\bf R})$.
From (\ref{force_diadic_form}) it can be shown that the
components of the displacements perpendicular to the plane 
do not contribute to the force in the harmonic approximation.
Each atom interacts with its three neighbors.
Denoting the masses of the two basis atoms by $m_1$ and $m_2$, and
using (\ref{force_diadic_form}) for the spring force acting on the atoms,
the following equations of motion are obtained for ${\bf u}({\bf R})$
and ${\bf v}({\bf R})$:
\begin{widetext}
\begin{subequations}
\begin{eqnarray}
m_1 \, \ddot{\bf u}({\bf R}) &=&
-D \, {\bf n_1} \otimes {\bf n_1} \,
\bigl[{\bf u}({\bf R}) - {\bf v}({\bf R})  \bigr]
-D \, {\bf n_2} \otimes {\bf n_2} \,
\bigl[{\bf u}({\bf R}) - {\bf v}({\bf R}-{\bf a}_1) \bigr]
-D \, {\bf n_3} \otimes {\bf n_3} \,
\bigl[{\bf u}({\bf R}) - {\bf v}({\bf R}-{\bf a}_2) \bigr], \\[1ex]
m_2 \, \ddot{\bf v}({\bf R}) &=&
-D \,{\bf n_1} \otimes {\bf n_1} \,
\bigl[{\bf v}({\bf R}) - {\bf u}({\bf R}) \bigr]
-D \, {\bf n_2} \otimes {\bf n_2} \,
\bigl[{\bf v}({\bf R}) - {\bf u}({\bf R}+{\bf a}_1)  \bigr]
-D \, {\bf n_3} \otimes {\bf n_3} \,
\bigl[{\bf v}({\bf R})- {\bf u}({\bf R}+{\bf a}_2)  \bigr].
\end{eqnarray}
\label{mozg:eq}
\end{subequations}
\end{widetext}
Eq.~(\ref{mozg:eq}) is an infinite set of equations for the displacements
${\bf u}({\bf R})$ and {\bf v}({\bf R}).
Following the traditional method, we seek a solution representing
a wave of angular frequency $\omega\left({\bf q}\right)$ and
wave vector ${\bf q}$:
\begin{subequations}
\begin{eqnarray}
{\bf u}({\bf R}) &=&
\frac{{\bf u}\left({\bf q} \right)}{\sqrt{m_1}} \,
 e^{i\omega\left({\bf q}\right) t +i{\bf qR}}, \\
{\bf v}({\bf R}) &=&
\frac{{\bf v}\left({\bf q} \right)}{\sqrt{m_2}}\,
e^{i\omega\left({\bf q}\right) t +i{\bf qR}},
\end{eqnarray}
\label{uv-fourier}
\end{subequations}
where the vectors ${\bf u}\left({\bf q}\right)$ and
${\bf v}\left({\bf q}\right)$ are to be determined,
and ${\bf q}$ is in the first Brillouin zone.
The usual Born-von K\'arm\'an periodic boundary conditions are
applied\cite{Kittel,Ziman,Ashcroft,Harrison,Jones,Lubensky}, i.e.\
${\bf u}({\bf R}+ N_i{\bf a}_i)= {\bf u}({\bf R})$ and
${\bf v}({\bf R}+ N_i{\bf a}_i)= {\bf v}({\bf R})$, where $N_1$ and $N_2$
are large integers whose product $N_1 N_2 = N$ is the total number of
primitive cells within the crystal.
The periodic boundary conditions restrict the allowed wave vectors
${\bf q}$ in Eq.\ (\ref{uv-fourier}) to the form:
\begin{equation}
{\bf q} = \frac{p_1}{N_1}{\bf b}_1 + \frac{p_2}{N_2}{\bf b}_2,
\label{k-vector}
\end{equation}
where $p_1$ and $p_2$ are integers, and the ${\bf b}_j$ are
the reciprocal lattice vectors defined
by ${\bf a}_i {\bf b}_j = 2\pi \delta_{ij}$, $i,j=1,2$.
It is convenient to choose $p_1, p_2$ such that ${\bf q}$ is
limited to the first Brillouin zone. For example, if $N_1$ and $N_2$ are
even integers (which will be irrelevant in the limit $N \rightarrow \infty$),
$-N_i/2 \leq p_i <  N_i/2$ ($i=1,2$).
More details on the concept of the Brillouin zone
can be found in many books on
solid-state physics\cite{Kittel,Ziman,Ashcroft,Harrison,Jones,Lubensky}.

Substitution of the solutions (\ref{uv-fourier}) into the
equations of motion (\ref{mozg:eq}) leads to an eigenvalue problem:
\begin{equation}
{\bf D}({\bf q})\, \begin{bmatrix}
{\bf u}\left({\bf q}\right) \\
{\bf v}\left({\bf q}\right)
\end{bmatrix} = \omega^2\left({\bf q}\right) \,
\begin{bmatrix}
{\bf u}\left({\bf q}\right) \\
{\bf v}\left({\bf q}\right)
\end{bmatrix} ,
\label{eigen_eq}
\end{equation}
where
\begin{subequations}
\begin{eqnarray}
{\bf D}({\bf q}) &=&
\begin{bmatrix}
{\bf D}_{11}({\bf q})  & {\bf D}_{12}({\bf q}) \\
{\bf D}_{21}({\bf q})  &  {\bf D}_{22}({\bf q})
\end{bmatrix} ,
\label{D_q:def} \\[2ex]
{\bf D}_{11}({\bf q}) & = & \frac {D}{m_1}
\sum_{j=1}^3 {\bf n}_j \otimes {\bf n}_j
 =  \frac {3D}{2m_1} {\bf E},
\label{D_11:def}  \\
{\bf D}_{22}({\bf q}) & = & \frac {D}{m_2}
\sum_{j=1}^3 {\bf n}_j \otimes {\bf n}_j
 =  \frac {3D}{2m_2} {\bf E},
\label{D_22:def} \\
{\bf D}_{12}({\bf q}) & = & -\frac {D}{\sqrt{m_1m_2}} e^{i{\bf q}{\bf d}_1}
\sum_{j=1}^3 {\bf n}_j \otimes {\bf n}_j \,
e^{-i{\bf q}{\bf d}_j},  \\
{\bf D}_{21}({\bf q}) & = & {\left [{\bf D}_{12}({\bf q})\right ]}^{*}.
\label{D_21:def}
\end{eqnarray}
\label{D_ij:def}
\end{subequations}
In the last equality of (\ref{D_11:def}) and (\ref{D_22:def})
we used Eq.~(\ref{n_jon_k:id}).
The matrix ${\bf D}({\bf q})$ is commonly called the 
\emph{dynamical matrix}. 
The eigenvectors of the dynamical matrix are the unknown vectors
${\bf u}\left({\bf q}\right)$ and ${\bf v}\left({\bf q}\right)$
of Eq.~(\ref{uv-fourier}).
In our case, ${\bf D}({\bf q})$ is a $4 \times 4$ matrix, while
the ${\bf D}_{ij}({\bf q})$ are  $2 \times 2$ matrices ($i,j=1,2$).
For a given ${\bf q}$ there are 4 eigenfrequencies and
4 corresponding, mutually orthogonal eigenvectors,
which give the amplitudes of the wave solutions (\ref{uv-fourier}).
In the literature these eigenvectors are called \emph{vibrational modes}.

The eigenfrequencies can be found by solving the following equation
for $\lambda ({\bf q}) = \omega^2\left({\bf q}\right)$:
\begin{equation}
{\rm det}\, \left [{\bf D}({\bf q}) -\lambda({\bf q})\, {\bf I} \right ] =0,
\label{BS:18b1}
\end{equation}
where ${\bf I}$ is the $4 \times 4$ unit matrix.
The $4 \times 4$ determinant can be evaluated directly; this is, 
however, rather tedious.
Below we present another approach, which allows us to find analytical
solutions for the eigenfrequencies.
Note that the dynamical matrix is hermitian (see Eq.~(\ref{D_21:def})),
therefore its eigenvalues are real.
The basic idea is that
${\bf D}({\bf q})$ is a block matrix, whose elements are the $2 \times 2$
matrices ${\bf D}_{ij}({\bf q})$, and ${\bf D}_{11}({\bf q})$ and
${\bf D}_{22}({\bf q})$ are diagonal.
This way, the original eigenvalue problem is reduced to finding
the eigenvalues of a $2 \times 2$ matrix.
This problem can be solved analytically.

\section{Analytic solution of the eigenvalue problem}
\label{algebra:seq}

In this section an analytic method is presented to determine the
eigenfrequencies and the vibrational modes of the honeycomb structure.
Since ${\bf D}_{11}({\bf q})$ and ${\bf D}_{22}({\bf q})$ are
diagonal matrices, one can eliminate
${\bf u}({\bf q})$ from Eq.~(\ref{eigen_eq}), and find
by using Eqs.~(\ref{D_11:def})-(\ref{D_22:def})
\begin{subequations}
\begin{eqnarray}
\!\!\!\!\!\!\!\!\!{\bf P}{\bf P}^* {\bf v}({\bf q})
&= & \overline{\lambda}({\bf q}) \,
{\bf v}({\bf q}),  \,\,\,\,\, \text{where}
\label{PP*-eigen}   \\
{\bf P} &=& \sum_{j=1}^3 {\bf n}_j \otimes {\bf n}_j \,
e^{i{\bf q}{\bf d}_j}, \label{P:def}  \\
\overline{\lambda}({\bf q}) &=&  \!\!\frac{m_1m_2}{D^2} \!\!
\left (\lambda({\bf q}) - \frac{3D}{2m_1}\right) \!\!
\left (\lambda({\bf q}) - \frac{3D}{2m_2}\right)\!\! .
\label{lambda-vonas}
\end{eqnarray}
\end{subequations}
Eq.~(\ref{PP*-eigen}) is another eigenvalue equation.
The matrix ${\bf P}{\bf P}^* $ can be rewritten as
\begin{eqnarray}
{\bf P}{\bf P}^* & =& \sum_{j,k=1}^3\, \left [
\left ({\bf n}_j \otimes {\bf n}_j\right )
\left ({\bf n}_k \otimes {\bf n}_k\right )\right ]
 e^{i{\bf q}\left ({\bf d}_j-{\bf d}_k \right )} \nonumber \\
 & = & \sum_{j,k=1}^3\, \left [
\left ({\bf n}_j \otimes {\bf n}_k\right )
\left ({\bf n}_j {\bf n}_k\right )\right ]
 e^{i{\bf q}\left ({\bf d}_j-{\bf d}_k \right )} \nonumber \\
& = & \sum_{j=1}^3 {\bf n}_j \otimes {\bf n}_j - \frac{1}{2}\,
\sum_{\substack{j,k=1 \\ j\ne k}}^3 {\bf n}_j \otimes {\bf n}_k
e^{i{\bf q}\left ({\bf d}_j-{\bf d}_k \right )}  \nonumber \\
& = & \frac{3}{2}\,\sum_{j=1}^3 {\bf n}_j \otimes {\bf n}_j -
\frac{1}{2}\, \sum_{j,k=1}^3 {\bf n}_j \otimes {\bf n}_k
e^{i{\bf q}\left ({\bf d}_j-{\bf d}_k \right )} \nonumber \\
& = & \frac{9}{4}\, {\bf E} - \frac{1}{2}\, {\bf f} \otimes {\bf f},
\label{BS:18:1l}
\end{eqnarray}
where
\begin{equation}
{\bf f}= \sum_{j=1}^3 {\bf n}_j
e^{i{\bf q}{\bf d}_j}.
\label{f:def}
\end{equation}
In the derivation, identities (\ref{direct-2:id}),
(\ref{n_jon_k:id}) and (\ref{n_jn_k:id}) have been used.
Note that ${\bf f}$ is a complex vector.

From (\ref{direct-1:id}) it is easy to see that ${\bf v}_1={\bf f}$ is
an eigenvector of the matrix ${\bf P}{\bf P}^* $:
\begin{equation}
{\bf P}{\bf P}^* {\bf v}_1 = {\bf P}{\bf P}^* {\bf f}
= \frac{9}{4}\, {\bf f} - \frac{1}{2}\, {\bf f}^2 \, {\bf f}
=\left(\frac{9}{4} -\frac{1}{2}\, {\bf f}^2 \right) {\bf v}_1 ,
\label{BS:18:1o}
\end{equation}
with eigenvalue
$\overline{\lambda}_1= 9/4 - 1/2 \, {\bf f}^2$.
Using (\ref{n_jn_k:id}), the dot product ${\bf f}^2$
can easily be evaluated:
\begin{eqnarray}
{\bf f}^2  &=&   \sum_{j,k=1}^3 ({\bf n}_j {\bf n}_k)
e^{-i{\bf q}{\bf d}_j+i{\bf q}{\bf d}_k }
=  3-\frac{1}{2}\, \sum_{\substack{j,k=1 \\ j\ne k}}^3
e^{i{\bf q}\left ({\bf d}_k-{\bf d}_j \right )}  \nonumber \\
&=&  3- \eta({\bf q}),
\label{BS:18:1q}
\end{eqnarray}
where
\begin{equation}
\eta({\bf q})=\cos {\bf q}{\bf a}_1+\cos {\bf q}{\bf a}_2 +
\cos {\bf q}({\bf a}_1-{\bf a}_2),
\label{eta:def}
\end{equation}
has been expressed in terms of the primitive translation vectors
${\bf a}_1$ and ${\bf a}_2$.
Thus, the eigenvalue
$\overline{\lambda}_1$ corresponding to the eigenvector ${\bf v}_1$ is
\begin{equation}
\overline{\lambda}_1=\frac{3+2\eta ({\bf q})}{4} .
\end{equation}

\begin{figure}[hbt]
\includegraphics[scale=1.2]{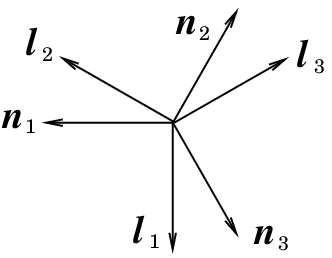}
\caption{The three vectors ${\bf l}_1,{\bf l}_2,{\bf l}_3$ are obtained by
a $90^{\circ}$-rotation from ${\bf n}_1,{\bf n}_2,{\bf n}_3$.
\label{lvectors:fig}}
\end{figure}

The other eigenvector must be perpendicular to ${\bf v}_1$.
It is useful to introduce three unit vectors,
${\bf l}_1,{\bf l}_2,{\bf l}_3$  (see Fig.~\ref{lvectors:fig}), such that
\begin{subequations}
\begin{eqnarray}
{\bf l}_j{\bf n}_j & = & 0,  \\
{\bf l}_j{\bf n}_k & = & -{\bf l}_k{\bf n}_j, \\
\absz {\,{\bf l}_j{\bf n}_k\,} & = & \frac{{\sqrt{3}}}{2}, \,\,\,
{\rm if} \, \, \, j \ne k .
\end{eqnarray}
\label{l_i:def}
\end{subequations}
Next, the vector 
\begin{equation}
{\bf g}= \sum_{j=1}^3 {\bf l}_j e^{-i{\bf q}{\bf d}_j}.
\label{g:def}
\end{equation}
is defined. Note the presence of the vectors ${\bf d}_j$ in
the exponent.
Then, from (\ref{l_i:def}) it is obvious that ${\bf g}$ is perpendicular
to the vector ${\bf f}$:
\begin{equation}
{\bf f}\,{\bf g}
= \sum_{j,k=1}^3 {\bf n}_j{\bf l}_k
e^{-i{\bf q}{\bf d}_j-i{\bf q}{\bf d}_k} = 0.
\label{BS:18:1y}
\end{equation}
Therefore,  the other eigenvector of ${\bf P}{\bf P}^* $ can be chosen
as ${\bf v}_2 = {\bf g}$,
and the corresponding eigenvalue $\overline{\lambda}_2$ can be found from
${\bf P}{\bf P}^*{\bf v}_2  = {\bf P}{\bf P}^*{\bf g}
=  \frac {9}{4}\, {\bf g} -
\frac {1}{2}\, {\bf f} \left({\bf f}\,{\bf g}\right)
= \frac {9}{4}\, {\bf v}_2.$
Thus, the second eigenvalue is
\begin{equation}
\overline{\lambda}_2 = \frac {9}{4}.
\label{lamv_2:eq}
\end{equation}
Note that this eigenvalue is independent of ${\bf q}$.
Finally, using (\ref{lambda-vonas}) one can obtain
the four eigenfrequencies of the dynamical matrix ${\bf D}({\bf q})$:
\begin{widetext}
\begin{subequations}
\begin{eqnarray}
\omega_{1,2}({\bf q}) &=& \sqrt{\lambda_{1,2}({\bf q})}  =
\sqrt{\frac {3}{4}\,D \left[\frac{1}{m_1}+\frac{1}{m_2} \pm
\sqrt{{\left(\frac{1}{m_1}-\frac{1}{m_2} \right)}^2 +
\frac{12+8\,\eta ({\bf q})}{9\, m_1m_2}}\,\right]},
\label{om_12:eq} \\[1ex]
\omega_{3}({\bf q}) & = &  \sqrt{\lambda_3({\bf q})}
=  0  \,\,\,\,\,\,  \text{and} \,\,\,\,\,\,
\omega_{4}({\bf q}) = \sqrt{\lambda_4({\bf q})}
= \sqrt{\frac {3}{2}D\left(\frac{1}{m_1}+\frac{1}{m_2} \right)},
\end{eqnarray}
\label{eigenfreki_veg}
\end{subequations}
\end{widetext}
where $\lambda_1$ and $\lambda_2$
($\pm$ signs, respectively)
correspond to $\overline{\lambda}_1$,
while $\lambda_3$ and $\lambda_4$ to $\overline{\lambda}_2$.
$\eta ({\bf q})$ is defined in (\ref{eta:def}).

To find ${\bf u}({\bf q})$ one can use
Eqs.~(\ref{eigen_eq}) and (\ref{D_11:def})-(\ref{D_22:def}),
and obtain
\begin{equation}
{\bf u}({\bf q})
= -\frac{D/\sqrt{m_1m_2}}{\lambda({\bf q}) - 3D/\left(2m_1\right)}\,
e^{i{\bf q}{\bf d}_1} {\bf P}^* {\bf v}({\bf q}).
\label{u:exp}
\end{equation}
First, ${\bf P}^* {\bf v}_1$ and ${\bf P}^* {\bf v}_2$ are calculated.
Using (\ref{direct-1:id}), (\ref{sum-n:id}) and (\ref{n_jn_k:id})
\begin{eqnarray}
{\bf P}^* {\bf v}_1  & = & {\bf P}^* {\bf f}
= \sum_{j=1}^3{\bf n}_j \otimes {\bf n}_j
e^{-i{\bf q}{\bf d}_j}\sum_{k=1}^3 {\bf n}_k
e^{i{\bf q}{\bf d}_k} \nonumber \\
& = & \sum_{j=1}^3{\bf n}_j - \frac{1}{2}\,
\sum_{\substack{j,k=1 \\ j\ne k}}^3{\bf n}_j
e^{i{\bf q}\left ({\bf d}_k-{\bf d}_j\right)}  \nonumber \\
& = & - \frac{1}{2}\, \sum_{k=1}^3 \,
e^{i{\bf q}{\bf d}_k}\sum_{j=1}^3{\bf n}_j
e^{-i{\bf q}{\bf d}_j}  =  c \, {\bf f}^*,
\label{Pstar-v_1:cal}
\end{eqnarray}
where
$c= - \frac{1}{2}\, \sum_{k=1}^3 \, e^{i{\bf q}{\bf d}_k}$ is a scalar
value depending on ${\bf q}$.
Similarly,
\begin{eqnarray}
 \!\!\!\! {\bf P}^* {\bf v}_2  & = & {\bf P}^* {\bf g}
= \sum_{j=1}^3 {\bf n}_j \otimes {\bf n}_j
e^{-i{\bf q}{\bf d}_j}\sum_{k=1}^3 {\bf l}_k
e^{-i{\bf q}{\bf d}_k} \nonumber \\
& = & \!\!\! -\frac{3}{2} \left ( {\bf l}_1 e^{i{\bf q}{\bf d}_1}
+{\bf l}_2 e^{i{\bf q}{\bf d}_2}
+{\bf l}_3 e^{i{\bf q}{\bf d}_3} \right )
\! = \! - \frac{3}{2}\, {\bf g}^*,
\label{Pstar-v_2:cal}
\end{eqnarray}
where we have made use of the identities
${\bf n}_2 -{\bf n}_1 = \sqrt{3}\, {\bf l}_3$,
${\bf n}_3 -{\bf n}_2 = \sqrt{3}\, {\bf l}_1$,
${\bf n}_1 -{\bf n}_3 = \sqrt{3}\, {\bf l}_2$ and
${\bf d}_1 + {\bf d}_2 + {\bf d}_3 = 0$.
Finally, using Eqs.~(\ref{u:exp})-(\ref{Pstar-v_2:cal})
the four eigenvectors are
\begin{subequations}
\begin{eqnarray}
\hspace{-1cm}& \omega_1  & \rightarrow
\begin{bmatrix}
-c\, C_1\left({\bf q}\right)\, {\bf f}^* \\
{\bf f}
\end{bmatrix}, \,\,\,
\omega_2 \rightarrow
\begin{bmatrix}
-c\, C_2\left({\bf q}\right)\, {\bf f}^* \\
{\bf f}
\end{bmatrix} \!\! , \\[3ex]
\hspace{-1cm}& \omega_3 & \rightarrow \!\!
\begin{bmatrix}
-\sqrt{\frac{m_1}{m_2}} \, {\bf g}^*  e^{i{\bf q}{\bf d}_1} \\
{\bf g}
\end{bmatrix},\ 
\omega_4 \rightarrow 
\begin{bmatrix}
\sqrt{\frac{m_2}{m_1}} \, {\bf g}^*  e^{i{\bf q}{\bf d}_1} \\
{\bf g}
\end{bmatrix} \!\! ,  \label{uv3uv4:eq}\\[3ex]
\hspace{-1cm}&& \text{where} \,\,\,  C_i\left({\bf q}\right)  = \!
\frac{D/\sqrt{m_1m_2}}{\lambda_i - 3D/\left(2m_1\right)} \,
 e^{i{\bf q}{\bf d}_1} ,
\end{eqnarray}
\end{subequations}
and  ${\bf f}$, ${\bf g}$, $\lambda_i$, and $c$ are
functions of {$\mathbf q$}
defined in Eqs.~(\ref{f:def}), (\ref{g:def}),
(\ref{eigenfreki_veg}) and after Eq.~(\ref{Pstar-v_1:cal}), respectively.
It can be shown that the four eigenvectors are mutually orthogonal.

It is also worth mentioning that the eigenmode corresponding to
$\omega_3({\bf q}) = 0$ is a special one, for example,
$\left[{\bf u}({\bf R}) - {\bf v}({\bf R})\right] {\bf n}_1 = 0$ 
for this mode.
This can easily be proved by using (\ref{uv-fourier}) and
the expression for the $\omega_3$ eigenmode given by (\ref{uv3uv4:eq}).
This means that neglecting the second order terms in the
displacements ${\bf u}({\bf R})$ and ${\bf v}({\bf R})$
all springs are {\em unstrained}, ie.\ their lengths remain $a$.
The atoms of the whole lattice move as if the spring system were a
universal joint. This movement of the atoms does not require any energy
(the eigenfrequency is zero), therefore there is no
resistance to a shear of the lattice. Consequently, the stability of
the lattice is lost, and the whole honeycomb structure becomes 
{\em unstable}.
In nature, besides the central interaction between nearest neighbors 
there exist  $n^{th}$ neighbor ($n=2,3,\cdots$) 
and lateral interactions as well\cite{Harrison,nanotube-book}.
In the latter case, the interatomic interaction energy depends
on the angle between two adjacent bonds.
Such interactions give rise to the so-called {\em bond-bending forces}.
These additional interactions stabilize the lattice 
and one must take them into account to calculate the vibrational modes 
more accurately. 
Nanotubes provide a good example for this\cite{nanotube-book}.

\section{Numerical results}
\label{res:seq}

\begin{figure}[hbt]
\includegraphics[scale=0.4]{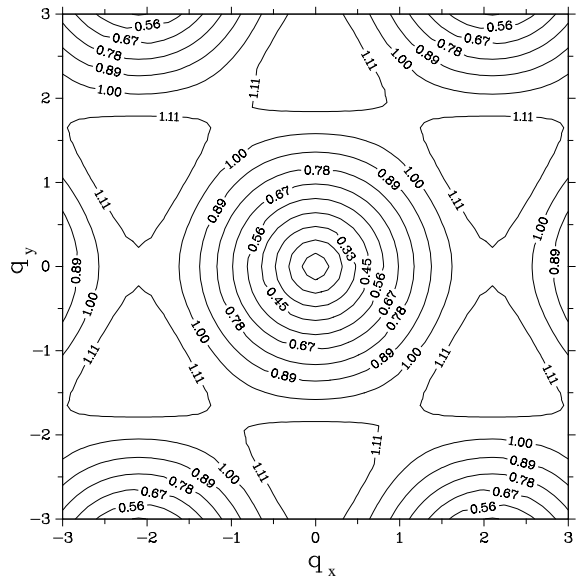}
\caption{The contour plot of $\omega_1({\bf q})$
(in units of $\sqrt{D/m_1}$) versus  ${\bf q}$ (in units of $1/a$)
for $m_1/m_2=2$. $q_x$ is parallel to the vector $-{\bf d}_1$.
\label{res-1:fig}}
\end{figure}

\begin{figure}[hbt]
\includegraphics[scale=0.4]{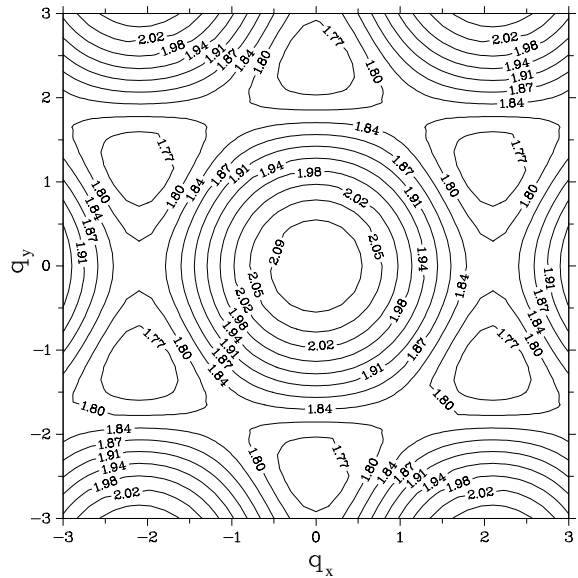}
\caption{The same contour plot as in Fig.~\ref{res-1:fig}  for
$\omega_2({\bf q})$.
\label{res-2:fig}}
\end{figure}

In this section a few numerical results for the eigenfrequencies
are presented. As it is seen from (\ref{eigenfreki_veg}) only
$\omega_1({\bf q})$ and  $\omega_2({\bf q})$ have nontrivial
${\bf q}$-dependence. In Figs.~\ref{res-1:fig} and \ref{res-2:fig}
the contour plot of $\omega_1({\bf q})$ and  $\omega_2({\bf q})$ are shown
for $m_1/m_2=2$. 
The vector ${\bf q}=(q_x,q_y)$ is taken in a Cartesian coordinate system in
which the $x$-axis is parallel to the direction of the vector $-{\bf d}_1$.
In these contour plots the lines show the constant eigenfrequencies 
in the parameter space  ${\bf q}=(q_x, q_y)$.  
It is seen from the figures that $\omega_1({\bf q})$ has
a deep minimum (zero) at ${\bf q}=0$, while $\omega_2({\bf q})$ has a
maximum there, equal to $\omega_4 = \sqrt{3D(1/m_1+1/m_2)/2}$ for arbitrary
mass ratio $m_1/m_2$.
The contour plot is also a useful representation of the vibrational modes 
to show the symmetries of the eigenfrequencies in the ${\bf q}$ space. 
One can see from Figs.~\ref{res-1:fig} and \ref{res-2:fig} that 
both $\omega_1({\bf q})$ and  $\omega_2({\bf q})$ 
have, eg.\ a $60^\circ$ rotation around
the axis which goes through the point ${\bf q} = 0$ 
and perpendicular to the $(q_x, q_y)$ plane.   

\begin{figure}[hbt]
\includegraphics[scale=0.4]{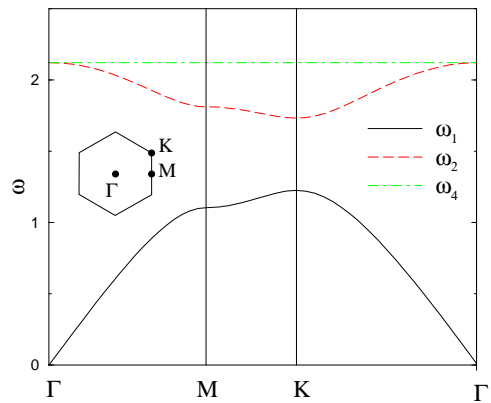}
\caption{The eigenfrequencies (in units of $\sqrt{D/m_1}$) 
$\omega_1({\bf q})$ (solid),
$\omega_2({\bf q})$ (dashed),
$\omega_4({\bf q})$ (dot-dashed) 
along lines between points $\Gamma$, M and K.
These points are in the first Brillouin zone shown in the inset. 
The mass ratio is $m_1/m_2=2$.
\label{res-3:fig}}
\end{figure}

In Fig.~\ref{res-3:fig} the eigenfrequencies $\omega_1({\bf q})$,
$\omega_2({\bf q})$ and  $\omega_4({\bf q})$ are plotted 
along the lines in the first Brillouin zone joining the points 
$\Gamma$, M and K shown in the inset of the figure.
In solid-state physics it is a common practice to plot the
eigenfrequencies along the lines between high symmetric points. 
Owing to the symmetry properties of the eigenfrequency in the 
${\bf q}$ space it is enough to calculate the eigenfrequencies 
only inside the area enlosed by these lines.  
In the same Cartesian coordinate system as in Fig.~\ref{res-1:fig} 
the points $\Gamma$, M and K are 
${\bf q}_{\rm {\scriptscriptstyle \Gamma}} = 0$, 
${\bf q}_{\rm {\scriptscriptstyle M}} 
= (\frac{2\pi}{3a},0)$ and 
${\bf q}_{\rm {\scriptscriptstyle K}} 
= (\frac{2\pi}{3a},\frac{2\pi}{3\sqrt{3}a})$, respectively. 
Here $a$ is the length of hexagons' side.  
Note that $\omega_4({\bf q})$ is independent of ${\bf q}$.

In the literature the vibrational mode corresponding to 
$\omega_1({\bf q})$ is called the {\em acoustical branch} 
because the eigenfrequency at small $|{\bf q}|$ has the same form as that
of the sound waves, namely $\omega=v|{\bf q}|$, where $v$ is the sound
velocity. Those vibrational modes for which the frequencies does not tend to
zero as $|{\bf q}|\rightarrow 0 $ form the {\em optical branch}.  
In our example $\omega_2({\bf q})$ and $\omega_4({\bf q})$ belong to the 
optical branch.  
It is clear from Fig.~\ref{res-3:fig} that at the point K there is a 
frequency gap between the acoustical and optical branch. 
This gap is equal to 
$\sqrt{\frac{3D}{2}}\left| 1/\sqrt{m_1}-1/\sqrt{m_2}\right|$.

\begin{figure}[hbt]
\includegraphics[scale=0.4]{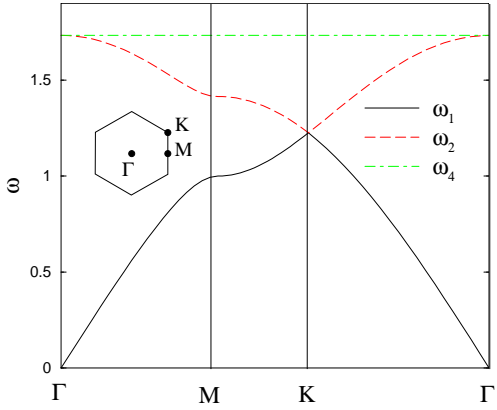}
\caption{The same plot as in Fig.~\ref{res-3:fig} for $m_1/m_2=1$. 
\label{res-4:fig}}
\end{figure}

We now take $m_1=m_2$ corresponding to a single graphite layer.
A similar plot as in Fig.~\ref{res-3:fig} is shown in Fig.~\ref{res-4:fig}
for this case. One can see that the gap disappears.  
These results are similar to those obtained from the recent 
accurate calculations for two modes of the isolated graphite 
layer\cite{Dubay-cikk}.

\section{Other lattice structures}
\label{3szog4szog:sec}

To demonstrate the effectivity of the general method developed 
in Sections \ref{force:seq} and \ref{motion:seq}, 
we now present further examples of lattice structures. 

\begin{figure}[hbt]
\includegraphics[scale=0.5]{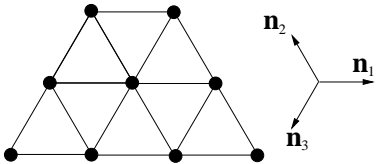}
\caption{Equilateral triangular lattice with nearest neighbor interactions. 
The straight solid lines correspond to unstrained springs of force constant
$D$ connecting nearest neighbor atoms in equilibrium state. 
The unit vectors ${\bf n}_1, {\bf n}_2$ and ${\bf n}_3$ correspond to the
directions of the unstrained springs of length $a$. 
\label{haromszog:fig}}
\end{figure}

In Fig.~\ref{haromszog:fig} an equilateral triangular lattice is shown. 
We assume only nearest neighbor interactions  repersented 
by springs of force constant $D$. 
One can easily write down the equations of motion for the displacements
of the atoms using 
Eq.~(\ref{force_diadic_form}) and it yields 
the $2 \times 2$ dynamical matrix: 
\begin{equation}
{\bf D}({\bf q})=\frac{2D}{m}\, \sum_{i=1}^{3} \, 
\left[1-\cos \left({\bf q} \, {\bf n}_i \, a\right) \right] 
\, {\bf n}_i  \otimes {\bf n}_i, 
\label{3szog-dinamic-diad:eq}  
\end{equation}
where the unit vectors ${\bf n}_i$ ($i=1,2,3$) are shown in
 Fig.~\ref{haromszog:fig}. 

\begin{figure}[hbt]
\includegraphics[scale=0.4]{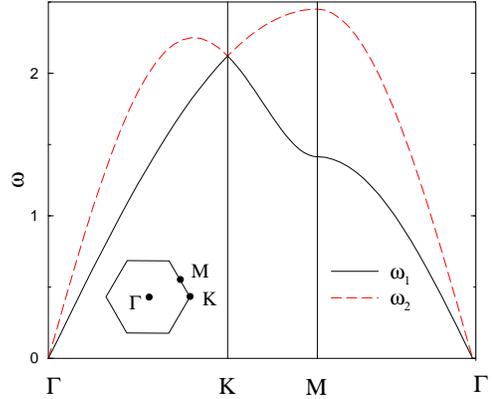}
\caption{For triangular lattice 
the eigenfrequencies (in units of $\sqrt{D/m}$) 
$\omega_1({\bf q})$ (solid) and $\omega_2({\bf q})$ (dashed) 
along lines between points $\Gamma$, K and M.
These points are in the first Brillouin zone shown in the inset.
\label{3szog-res:fig}}
\end{figure}

The eigenfrequencies are again the square-root of the eigenvalues of the
dynamical matrix as in (\ref{BS:18b1}). 
From a simple calculation one finds
\begin{equation}
\omega_{1,2}({\bf q})= 
\sqrt{\frac{D_{11}+ D_{22}\pm 
\sqrt{{\left(D_{11}-D_{22}\right)}^2 + 4 D_{12}^2}}{2}},
\label{sajatertek_22:eq}
\end{equation}
where $D_{ij}$ is the $ij$ matrix element of the $2 \times 2$ 
dynamical matrix ${\bf D}({\bf q})$ for a given wave vector ${\bf q}$. 
In Fig.~\ref{3szog-res:fig} the eigenfrequencies $\omega_1({\bf q})$ 
and $\omega_2({\bf q})$ are plotted 
along the lines in the first Brillouin zone joining the points 
$\Gamma$, K and M shown in the figure. 
In the Cartesian coordinate system in which $q_x$ is parallel to the unit
 vector ${\bf n}_1$,  
the points $\Gamma$, X and M are 
${\bf q}_{\rm {\scriptscriptstyle \Gamma}} = 0$, 
${\bf q}_{\rm {\scriptscriptstyle K}} 
= (\frac{4\pi}{3a},0)$ and 
${\bf q}_{\rm {\scriptscriptstyle M}} 
= (\frac{\pi}{a},\frac{\pi}{\sqrt{3}a})$, respectively.    

\begin{figure}[hbt]
\includegraphics[scale=0.4]{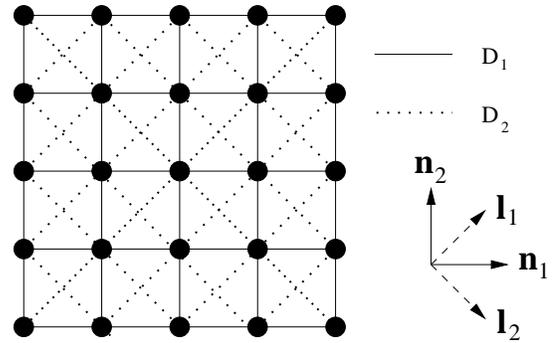}
\caption{Square lattice with first and second nearest interactions. 
The straight lines correspond to unstrained springs of force constant
$D_1$ (solid lines) and $D_2$ (dashed lines) 
connecting first and second nearest neighbor atoms in equilibrium state.
The distance between nearest neighbors is $a$.
The unit vectors ${\bf n}_1, {\bf n}_2, {\bf l}_1$ and ${\bf l}_2$ 
correspond to the directions of the unstrained springs. 
\label{fnegyzet:fig}}
\end{figure}

\begin{figure}[hbt]
\includegraphics[scale=0.4]{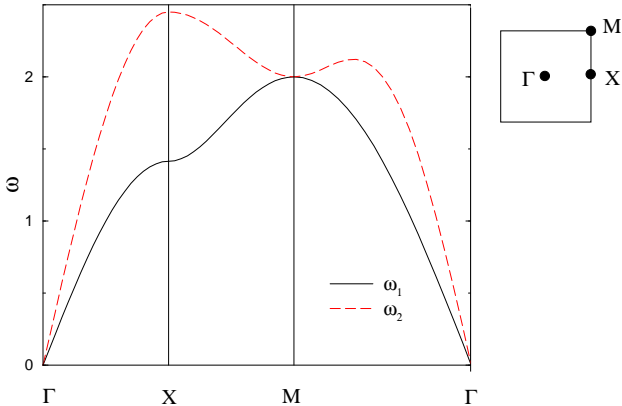}
\caption{For square lattice 
the eigenfrequencies (in units of $\sqrt{D_1/m}$) 
$\omega_1({\bf q})$ (solid) and $\omega_2({\bf q})$ (dashed) 
along lines between points $\Gamma$, X and M.
These points are in the first Brillouin zone shown on the right of the
figure.  
The ratio of the force constants is $D_1/D_2=2$.
\label{negyzet-res:fig}}
\end{figure}

In Fig.~\ref{fnegyzet:fig} a square lattice is shown. 
We now assume first and second nearest neighbor interactions (solid and
dashed lines in the figure) represented 
by springs of force constant $D_1$ and $D_2$, respectively. 
Similar way as before, one finds that the $2 \times 2$ dynamical matrix is    
\begin{eqnarray}
{\bf D}({\bf q}) &=& 
\frac{2 D_1}{m}\, \sum_{i=1}^2 \, 
\left[1-\cos \left({\bf q} \, {\bf n}_i \, a\right) \right] \, 
{\bf n}_i \otimes {\bf n}_i 
\nonumber \\
&+& \frac{2 D_2}{m}\, 
\sum_{i=1}^2\, 
\left[1-\cos \left({\bf q} \, {\bf l}_i \sqrt{2}\, a\right) \right] \, 
{\bf l}_i \otimes {\bf l}_i ,
\label{negyzet-D-diad:eq}
\end{eqnarray}
where the unit vectors ${\bf n}_i$ and ${\bf l}_i$ ($i=1,2$) are shown in
 Fig.~\ref{fnegyzet:fig}. 
Finally, the two eigenfrequencies can be obtained again from 
Eq.~(\ref{sajatertek_22:eq}).
In Fig.~\ref{negyzet-res:fig} the eigenfrequencies $\omega_1({\bf q})$ 
and $\omega_2({\bf q})$ are plotted 
along the lines in the first Brillouin zone joining the points 
$\Gamma$, X and M shown in the figure. 
In the Cartesian coordinate system in which $q_x$ is parallel to the unit
 vector ${\bf n}_1$,  
the points $\Gamma$, X and M are 
${\bf q}_{\rm {\scriptscriptstyle \Gamma}} = 0$, 
${\bf q}_{\rm {\scriptscriptstyle X}} 
= (\frac{\pi}{a},0)$ and 
${\bf q}_{\rm {\scriptscriptstyle M}} 
= (\frac{\pi}{a},\frac{\pi}{a})$, respectively.     

\section{Conclusions}
\label{veg}

We first calculated the vibrational modes of the honeycomb lattice in
the harmonic approximation. It was assumed that nearest
neighbor atoms are connected by ideal springs.
Using the direct product of vectors we derived a formula for
the spring force acting on an atom.
The equations of motion for the atoms were then derived and the resulting
dynamical matrix was given explicitly.
The vibrational frequencies and modes were determined from the eigenvalue
problem of the dynamical matrix by analytic methods.
Our work may provide a starting point to the studies of the more
complicated lattice dynamics of nanotubes.
Our general approach is also applied to study the lattice dynamics of the
equilateral triangular and the square lattices. 
In the latter case, to investigate more complicated structures, not only
the first but the second nearest neighbor interactions are also included.

Finally, we mention some problems.

\begin{figure}[hbt]
\includegraphics[scale=0.5]{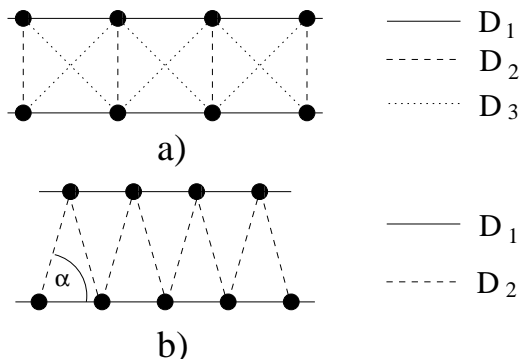}
\caption{Square (a) and triangular (b) ladder lattices.
Different types of lines represent different spring constants.
In the triangular ladder all springs of force constant $D_2$
have the same unstrained length.
\label{letrak:fig}}
\end{figure}

\begin{itemize}

\item
What is the first Brillouin zone for the square, triangular and 
honeycomb lattice structures discussed before?
Study the symmetry properties of the eigenfrequencies
$\omega_1({\bf q})$ and $\omega_2({\bf q})$ in the first Brillouin zone.

\item
Study the long wavelength limit, that is when $|{\bf q}|a \ll 1$ for 
the square, triangular and honeycomb lattice structures.
What are the eigenmodes in this case?
Determine the sound velocity.

\item
In Fig.~\ref{letrak:fig} a and b two ladder-type lattices are shown.
Using the method developed in Sections \ref{force:seq} and \ref{motion:seq}
find the vibrational frequencies and modes for these lattices.
Determine how the results depend on $\alpha$
in Fig.~\ref{letrak:fig} b.

\end{itemize}

\acknowledgments

We thank A.~Pir{\'o}th, J.~K\"urti and L.~Mih\'aly for helpful discussions.
This work was supported by the Hungarian  Science Foundation OTKA  TO34832.




\end{document}